\documentclass[
	prl, aps,
	superscriptaddress,
	twocolumn,
]{revtex4-1}

\usepackage{amsmath}
\usepackage{amssymb}
\usepackage{graphicx}
\usepackage{units}
\usepackage{upgreek}

\usepackage{xcolor}
\usepackage{dsfont} % double stroke font
\usepackage{xspace}

\newcommand{\ie}{\mbox{i.\hspace{0.125em}e.}\@\xspace}
\newcommand{\eg}{\mbox{e.\hspace{0.125em}g.}\@\xspace}
\definecolor{plot1}{RGB}{6,115,183}
\definecolor{plot2}{RGB}{255,118,0}
\definecolor{plot3}{RGB}{0,169,25}
\definecolor{plot4}{RGB}{230,0,28}
\definecolor{plot5}{RGB}{0,0,0}
\definecolor{plotBlue}{RGB}{6,115,183}
\definecolor{plotOrange}{RGB}{255,118,0}
\definecolor{plotGreen}{RGB}{0,169,25}
\definecolor{plotRed}{RGB}{230,0,28}

\newcommand{\Eqref}[1]{\eqref{#1}}

\newcommand{\figref}[1]{\mbox{Fig.\hspace{0.25em}\ref{#1}}}

\newcommand{\tabref}[1]{\mbox{Tab.\hspace{0.25em}\ref{#1}}}

\newcommand{\diff}{\text{d}}

\newcommand{\Nabla}{\vect\nabla}

\DeclareFontFamily{U}{mathx}{\hyphenchar\font45}
\DeclareFontShape{U}{mathx}{m}{n}{<-> mathx10}{}
\DeclareSymbolFont{mathx}{U}{mathx}{m}{n}
\DeclareMathAccent{\widebar}{0}{mathx}{"73}
\newcommand{\vect}{\boldsymbol}

\def\barroman#1{\sbox0{#1}\dimen0=\dimexpr\wd0+1pt\relax
  \makebox[\dimen0]{\rlap{\vrule width\dimen0 height 0.06ex depth 0.06ex}%
    \rlap{\vrule width\dimen0 height\dimexpr\ht0+0.03ex\relax 
            depth\dimexpr-\ht0+0.09ex\relax}%
    \kern.5pt#1\kern.5pt}}

\newcommand{\Reyn}{\operatorname{\mathit{R\kern-.08em e}}}
\newcommand{\Pran}{\operatorname{\mathit{P\kern-.08em r}}}
\newcommand{\Pecl}{\operatorname{\mathit{P\kern-.08em e}}}
\newcommand{\Nuss}{\operatorname{\mathit{N\kern-.2em u}}}
\newcommand{\Sher}{\operatorname{\mathit{S\kern-.08em h}}}
\newcommand{\Schm}{\operatorname{\mathit{S\kern-.08em c}}}
\newcommand{\Grae}{\operatorname{\mathit{G\kern-.08em z}}}
\newcommand{\Dean}{\operatorname{\mathit{D\kern-.08em e}}}
\newcommand{\Worm}{\operatorname{\mathit{W\kern-.24em o}}}

\newcommand{\dP}{P}

\newcommand{\Rh}{R_{\rm h}}
\newcommand{\Len}{\lambda}

\newcommand{\cambient}{c_\mathrm a}
\newcommand{\cbody}{c_\mathrm b}
\newcommand{\cwall}{c_\mathrm w}
\newcommand{\Dheat}{D_\mathrm h}
\newcommand{\Dvapor}{D_\mathrm w}
\newcommand{\Vt}{V_\mathrm t}

\newcommand{\viscKin}{\nu}
\newcommand{\viscDyn}{\eta}
\newcommand{\resist}{K}
\newcommand{\exchange}{\mathcal E}
\newcommand{\exchangeIn}{\exchange_\mathrm{in}}
\newcommand{\exchangeEx}{\exchange_\mathrm{ex}}
\newcommand{\volFlux}{Q}
\newcommand{\rateResp}{f}

\renewcommand{\figref}[1]{Fig.\nolinebreak[4]\hspace{0.25em}\nolinebreak[4]\ref{#1}}
\renewcommand{\tabref}[1]{Table~\ref{#1}}

\begin{document}

\title{Physical and geometric constraints explain the labyrinth-like shape of the nasal cavity}

\author{David Zwicker}
\affiliation{John A. Paulson School of Engineering and Applied Sciences, Harvard University, Cambridge, MA 02138, USA}
\affiliation{Kavli Institute for Bionano Science and Technology, Harvard University, Cambridge, MA 02138, USA}

\author{Rodolfo Ostilla-M\'{o}nico}
\affiliation{John A. Paulson School of Engineering and Applied Sciences, Harvard University, Cambridge, MA 02138, USA}
\affiliation{Kavli Institute for Bionano Science and Technology, Harvard University, Cambridge, MA 02138, USA}

\author{Daniel E. Lieberman}
\affiliation{Department of Human Evolutionary Biology, Harvard University, 11 Divinity Avenue, Cambridge, MA 02138, USA}

\author{Michael P. Brenner}
\affiliation{John A. Paulson School of Engineering and Applied Sciences, Harvard University, Cambridge, MA 02138, USA}
\affiliation{Kavli Institute for Bionano Science and Technology, Harvard University, Cambridge, MA 02138, USA}

\begin{abstract}
The nasal cavity is a vital component of the respiratory system that heats and humidifies  inhaled air in all vertebrates.
Despite this common function, the shapes of nasal cavities vary widely across animals.
To understand this variability, we here connect nasal geometry to its function by theoretically studying the airflow and the associated scalar exchange that describes heating and humidification.
We find that optimal geometries, which have minimal resistance for a given exchange efficiency, have a constant gap width between their side walls, but their overall shape is restricted only by the geometry of the head.
Our theory explains the geometric variations of natural nasal cavities quantitatively and we hypothesize that the trade-off between high exchange efficiency and low resistance to airflow is the main driving force shaping the nasal cavity.
Our model further explains why humans, whose nasal cavities evolved to be smaller than expected for their size, become obligate oral breathers in aerobically challenging situations.
\end{abstract}

\maketitle

\begin{figure}
	\centerline{
		\includegraphics[width=\columnwidth]{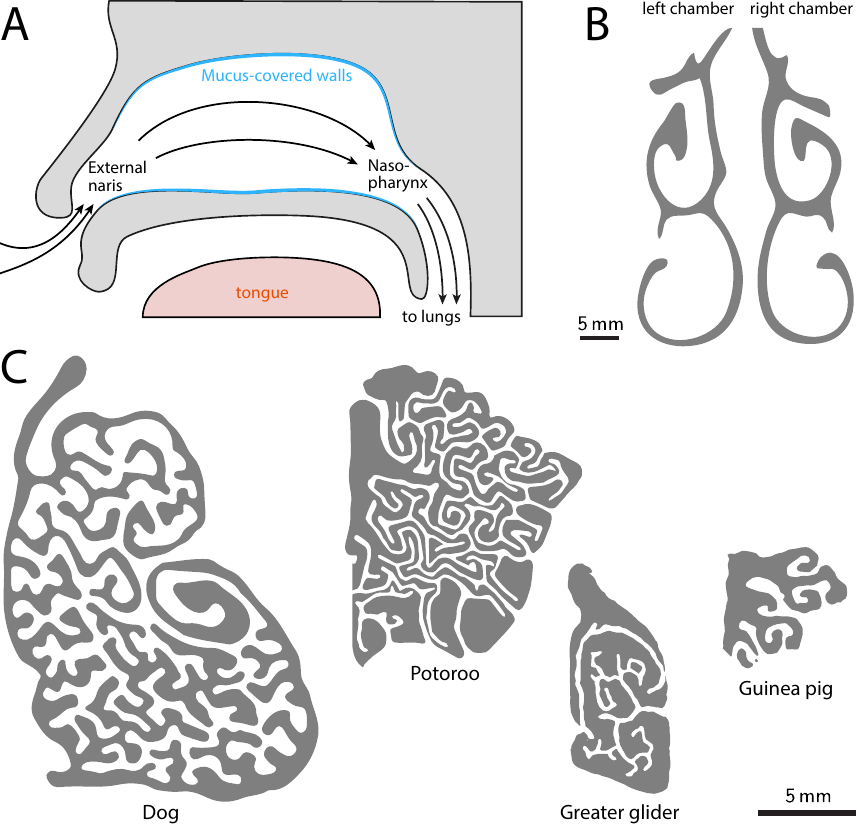}
	}
	\caption{
	Schematic cross-sections of nasal cavities.
	(A) Sagittal cross-section showing how air flows though the nasal cavity during inhalation.
	(B) Coronal cross-section of a human nasal cavity showing the complex shape of the two air-filled nasal chambers (gray region). The airflow is perpendicular to the plane.
	(C) Coronal cross-sections of the right nasal chambers of mammals (sorted from left to right by decreasing body weight):
	\textit{Canis lupus familiaris} (Dog, \cite{Craven2009}),
	\textit{Potorous tridactylus} (Long-nosed potoroo, \cite{Macrini2012}),
	\textit{Petauroides volans} (Great Glider, \cite{Macrini2012}), and
	\textit{Cavia porcellus} (Guinea pig, \cite{Digimorph}).
	Shapes have been reconstructed from CT scans obtained from the cited sources.
	Scale bars in (B) and (C) equal \unit[5]{mm}.
	}
	\label{fig:schematic}
\end{figure}

The nose not only allows us to smell but it also humidifies, heats, and cleans  inhaled air before it reaches the lungs.
All these vital tasks depend critically on nasal airflow, which is driven by the pressure difference created by the lungs and depends on the complex geometry of the nasal cavity. 
Nasal geometries vary considerably among vertebrates in general~\cite{Witmer1995} and among mammals in particular \cite{Negus1954, Macrini2012, Maier2014}
, ranging from the complex labyrinth-like internal nasal cavity of dogs to the unique structure of humans that combines relatively simple geometry in a short internal nasal cavity with an additional external nasal vestibule, see \figref{fig:schematic}.
These qualitative differences in nasal geometry were likely selected as adaptations to different functional requirements, but how the geometry of the nose influences the airflow and thus the function of the nose is a long-standing unsolved problem.

The nasal cavity is a complex, air-filled space that connects the two nostrils with the throat, see \figref{fig:schematic}A.
All mammals have an internal nasal cavity, but humans are unique in having an additional external vestibule with inferiorly oriented nostrils~\cite{Lieberman2011}.
The two sides of the cavity are separated by the nasal septum and merge only behind the posterior nasal cavity (choanae) that separate the nasal cavity from the pharynx.
Each side can be further divided into the main pathway (turbinates) and large side chambers (sinuses).
The walls of the nasal cavity are covered by a highly vascularized bed of epithelial tissue overlain by a $\unit[10]{\upmu m}$ thick layer of mucus, which is slowly propelled backwards by cilia~\cite{Quraishi1998}.
The mucus consists mainly of water and thus humidifies inhaled air.
Additionally, the nasal epithelium warms the air and absorbs airborne particles, like odorants.

We here study how the geometry of the nasal cavity influences the airflow and the associated processes of heating and humidifying the inhaled air.
Generally, we expect that a narrower geometry improves the efficiency of heating and humidification at the expense of greater resistance to airflow.
Since this trade-off likely plays an important role in shaping nasal cavities, we ask which shape has the lowest resistance to airflow for a given conditioning of the inhaled air.
Here, we have to take into account geometric constraints imposed by the shape of the head that determines the length of the nasal cavity, its cross-sectional area, and generally the shape of the space that it occupies.
To tackle this complex problem, we first analyze the properties of airflow in general nasal cavities, then investigate the resistance and conditioning of simple shapes quantitatively, and finally discuss the influence of complex geometric constraints.

\section*{Results}

\subsection{The flow in the nasal cavity is laminar}

It has been suggested that the flow in the nasal cavity is turbulent~\cite{Churchill2004, Clement2005}, since the speeds are high and nasal geometry is complex.
Indeed, turbulence can easily be induced in the surrounding air by exhaling heavily, as apparent on a cold winter day.
Inside the nose, turbulent flow would induce additional mixing that improves the heating and humidification of the inhaled air~\cite{Bergman2011}, but it also implies a larger resistance to flow.
It is thus unclear whether turbulence would be beneficial.

To see whether turbulence occurs inside the nasal cavity, we first estimate the mean speed~$\bar u$ using experimentally determined scaling relations of respiratory quantities with body mass, see \tabref{tab:scalings}.
In particular, we combine the volumetric flux~$Q$, the volume~$V$ of the nasal cavity, and the length of the skull as a proxy for the length~$L$ of the cavity, to obtain $\bar u = QL/V \approx \unitfrac[0.3]{m}{s} \cdot (M/\unit{kg})^{0.15 \pm 0.06}$.
Since $\bar u$ is much smaller than the speed of sound, the flow is incompressible.
For the flow to be turbulent, inertia has to dominate viscous dampening.
The ratio of these two effects is quantified by the Reynolds number~$\Reyn=\bar u \Rh / \viscKin$, where $\viscKin \approx \unitfrac[1.5 \cdot 10^{-5}]{m^2}{s}$ is the kinematic viscosity of air~\cite{Gates2012} and $\Rh$ is the characteristic linear dimension of the flow.
In complex geometries, $\Rh$ is given by the hydraulic radius $\Rh=2V/S$, where $V$ and $S$ are the volume and surface area of the geometry, respectively.
Using the scalings given in \tabref{tab:scalings}, we find $\Reyn = 2LQ / (\nu S) \approx 70 \cdot (M/\unit{kg})^{0.36 \pm 0.06}$, so  $\Reyn$ increases with body mass.
Typically, flows are turbulent when $\Reyn$ is about $1000$, with the precise transition depending weakly on the flow geometry; for example, parallel plates~\cite{Sano2016} are turbulent at a lower $\Reyn$ than  pipes~\cite{Eckhardt2007}. Regardless, this shows that
 turbulence occurs only in animals heavier than about $\unit[2000]{kg}$.
Indeed, numerical simulations have shown that the narrow geometry of the nasal cavity prevents the development of turbulence in humans~\cite{Shi2006}, dogs~\cite{Craven2009}, and rats~\cite{Zhao2006}.

\begin{table}
 \caption{Scaling of respiratory quantities~$y$ with body mass~$M$ given in units of $\unit{kg}$, $y=a (M/\unit{kg})^b$.
 }
 \label{tab:scalings}
  \centering\small
  \begin{tabular*}{\hsize}{@{\extracolsep{\fill}}llrr}

    Quantity~$y$ & Pre-factor~$a$ & Exponent~$b$ & Source \\ 
    \hline
    Skull length~$L$ & $\unit[7]{cm}$ & $0.32 \pm 0.01$ & \cite{Green2012} \\ 
    Surface area~$S$ & $\unit[19]{cm^2}$ & $0.74 \pm 0.03$ & \cite{Green2012} \\ 
    Nasal volume~$V$ & $\unit[2.9]{cm^3}$ & $0.96 \pm 0.03$ & \cite{Green2012} \\
    Tidal volume~$\Vt$ & $\unit[7.7]{cm^3}$ & $1.04 \pm 0.01 $ & \cite{Stahl1967} \\
    Respiratory rate~$\rateResp$ & $\unit[0.89]{s^{-1}}$ & $-0.26 \pm 0.01$ & \cite{Stahl1967} \\
  	\hline
	Cross-section~$A$ &$\unit[0.4]{cm^2}$ & $0.63 \pm 0.04$ &  $A = V/L$ \\ 
    Volumetric flux~$Q$ &$\unitfrac[14]{cm^3}{s}$ & $0.78 \pm 0.02$ & $\volFlux=2\Vt\rateResp$ \\
    \hline
    \end{tabular*}
\end{table}

Another ubiquitous feature of nasal cavity flow is the oscillatory motion caused by natural breathing, which generally induces additional resistance and also limits the humidification and heating of the air.
However, this is only important when the characteristic length~$\Rh$ is smaller than the length $(\viscKin/\rateResp)^{1/2}$ associated with the frequency~$f$ of the flow~\cite{Womersley1955}.
Using the scalings from \tabref{tab:scalings}, we find $\Rh (\rateResp/\viscKin)^{1/2} \approx 0.7 \cdot (M/\unit{kg})^{0.09 \pm 0.07}$, which depends only weakly on the body mass~$M$, so the effect of pulsatility is similar for all animals.
We show in the SI that the resistance increases by about $\unit[50]{\%}$ compared to steady flow, since $\Rh$ is comparable to $(\rateResp/\viscKin)^{1/2}$.
Note that if the frequency~$\rateResp$ were much higher, the resistance would increase strongly, while a lower frequency would imply a larger tidal volume and thus a larger lung.
The associated trade-off might set the respiratory rate, but since the pulsatility affects all animals similarly, we can simply analyze steady flow here.

\subsection{The optimal nasal cavity has a constant gap width}
We seek the geometry of the nasal cavity with the lowest resistance to airflow for a given efficiency of heating and humidifying the air under the constraint of a given volumetric flux, length, and cross-sectional area.
We thus need to calculate the dependence of the airflow and its physical properties on the  geometry of the nasal cavity.
Since nasal cavities are typically straight, we first focus on varying the cross-sectional shape.

The flow through the cavity is driven by a pressure difference~$P$ generated by the lungs.
Since the flow in a straight nasal cavity is  laminar, stationary, and incompressible, the velocity field only has a component~$u$ in the axial direction, which obeys the Poisson equation
\begin{equation}
	\Nabla^2_\perp u(x, y) = -\frac{P}{\viscDyn L}
	\label{eqn:flow}
\end{equation}
with $u=0$ at the walls, see SI. 
Here, $\Nabla^2_\perp$ denotes the Laplacian in the cross-sectional plane, $L$ is the length of the cavity, and $\viscDyn \approx \unit[1.8 \cdot 10^{-5}]{Pa \,s}$ is the dynamic viscosity of air~\cite{Gates2012}.
Solving \Eqref{eqn:flow} for $u$, we calculate the volumetric flux $Q = \int u \, \diff x \diff y$, which scales with $P$.
The resistance~$\resist=\dP/Q$ is then independent of $Q$ and can be expressed as
\begin{equation}
	\resist = C_\resist \cdot \frac{\viscDyn L}{A^2}
	\label{eqn:resist_scaling}
	\;,
\end{equation}
where $C_\resist$ is a non-dimensional parameter that depends only on the cross-sectional shape, see SI.
We can thus quantify the influence of the cross-sectional shape on the airflow by simply studying its effect on $C_\resist$.

The heating and humidification properties of the nasal cavity can be quantified by the change in temperature and concentration of water vapor in the air after it flowed through the cavity.
Both quantities can be described as a passive scalar~$c$ that is transported with the flow, diffuses, and is exchanged with the walls of the cavity.
In a stationary state, the scalar~$c$ thus obeys
\begin{equation}
	0 = D\Nabla^2 c - \partial_z(u c)
	\;,
\end{equation}
where $D$ denotes the diffusivity and boundary conditions are imposed by the epithelial tissue.
For simplicity, we first consider the case where the boundary is kept at body temperature and maximal humidity, which implies a constant scalar value $\cbody$ at the walls.
While flowing through the cavity, the cross-sectionally averaged scalar~$\bar c(z)$ will thus change from its ambient value~$\bar c(0) = \cambient$ to approach $\cbody$.
The extent of this process at the end of the cavity is quantified by the scalar exchange efficiency
\begin{equation}
	\exchange = \ln\left(\frac{\bar c(0) - \cbody}{\bar c(L) - \cbody}\right)
	\label{eqn:exchange_definition}
	\;,
\end{equation}
which is larger the closer $\bar c(L)$ gets to $\cbody$.
We show in the SI that $\bar c(z)$ can be expressed as $\bar c(z) = \cbody + \sum_n a_n e^{-z/\Len_n}$, where $\Len_n$ are length scales that follow from the generalized eigenvalue problem

\begin{equation}
	- D  \Nabla^2_\perp c_n(x, y)  = \Len_n^{-1}  u(x, y) c_n(x, y)
	\label{eqn:scalar_eigensystem}
\end{equation}
and the coefficients $a_n$ can be determined from the initial value~$\bar c(0)$ at the inlet.
Note that we here neglected the small axial diffusion term since it is dominated by advection ($\bar u L \gg D$).
Moreover, in the simple case of a long cavity, entrance effects can be neglected and only the mode with the largest~$\Len_n$ contributes to~$\exchange$.
In this case, we have 
\begin{equation}
	\exchange = C_\exchange \cdot \frac{D L}{\volFlux}
	\;,
	\label{eqn:exchange_scaling}
\end{equation}
where $C_\exchange= Q/(\lambda_1 D)$ is a non-dimensional factor associated with the largest~$\Len_n$.
We show in the SI that $C_\exchange$ depends only on the cross-sectional shape and captures how the shape affects the scalar exchange.

\begin{figure}
	\centerline{
		\includegraphics[width=\columnwidth]{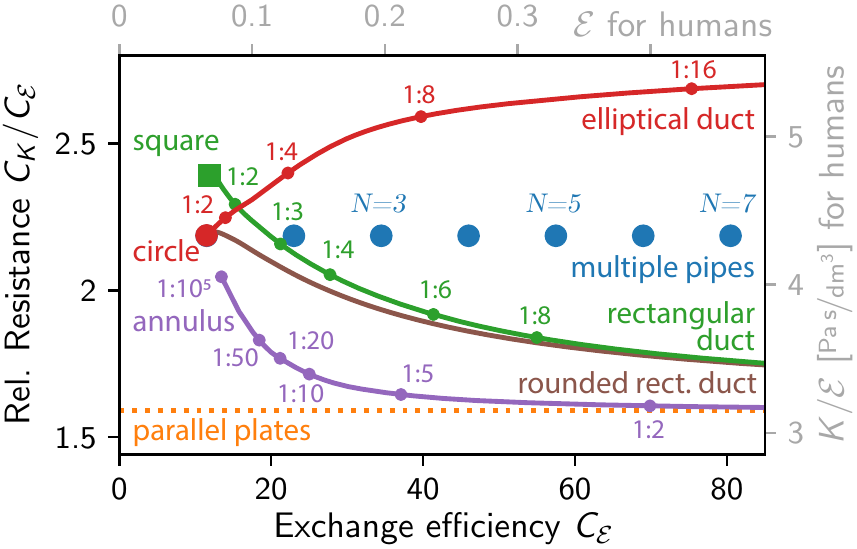}
	}
	\caption{
	Comparison of the relative resistance~$\resist / \exchange$ for different geometries as a function of a prescribed exchange efficiency~$\exchange$.
	Only the pre-factors $C_\resist$ and $C_\exchange$ are shown, but the top and right axes display the respective values for a standard human nose with parameters given in \tabref{tab:parameters}.
	The considered shapes are
	a single circular pipe (large red circle),
	a square duct (green square),
	elliptical ducts (red line, ratio of half-axes indicated at support points),
	rectangular ducts (green line, ratio of sides indicated at support points),
	rectangular ducts with rounded corners (brown line),
	$N$ circular pipes in parallel (blue dots), and
	annulus-shaped ducts (violet, ratio of inner to outer radius indicated at support points).
	The resistance of the proper cross-sections are bounded by the value for parallel plates (orange dotted line).
	Solid lines indicate results from quadratic interpolations between the support points (dots) were values were calculated numerically and cross-checked with the literature~\cite{Shah1971}.
	\label{fig:geometries}
	}
\end{figure}

To find the geometry that has minimal resistance~$\resist$ for a given exchange efficiency~$\exchange$, we determine the respective pre-factors $C_\resist$ and $C_\exchange$ for several simple shapes.
\figref{fig:geometries} shows that narrower channels generally have lower resistance for a given exchange efficiency.
In particular, shapes with aspect ratios close to unity, like the circle and the square, perform badly.
However, a large aspect ratio is not necessarily advantageous as elliptical shapes (red line) are far worse than rectangular ones (green line), particularly at large aspect ratios.
Instead, a constant gap width seems to be important, since this is the geometric feature shared between the optimal shapes.
Additionally, sharp corners seem to be detrimental, since the resistance decreases when the corners of a rectangular duct are rounded (brown line).
Taken together, we expect that the optimal nasal cavity exhibits an elongated, rounded cross-sectional shape with a constant gap width.

The optimal gap width can be estimated from the asymptotic case of two parallel plates, which provides a lower bound for the achievable resistance, see \figref{fig:geometries}.
This geometry corresponds to a rectangular duct where the two small sides are replaced by unphysical periodic boundary conditions, so the cross-sectional area~$A$ is still well defined.
The pre-factor for the scalar exchange in this geometry is $C_\exchange = 7.54 \, A \ell^{-2}$, where $\ell$ is the plate separation.
Using \Eqref{eqn:exchange_scaling}, we can then solve for the $\ell$ that results in a given scalar exchange efficiency $\exchange$,
\begin{equation}
	\ell = 2.75 \sqrt{\frac{D\tau}{\exchange}}
	\label{eqn:gap_width}
	\;,
\end{equation}
where $\tau = LA/Q$ is the time it takes air to cross the whole cavity.
The gap width must thus be similar to the typical distance~$(D\tau)^{1/2}$ the scalar diffuses while passing through the cavity~\cite{Squires2008}.

The result of our theoretical considerations is twofold:
First, we qualitatively predict that natural selection should favor nasal cavities that mimic a (rounded) rectangular channel with high aspect ratio, \ie, the separation between the walls should be approximately constant everywhere in the nasal cavity.
Second, we quantitatively predict this gap width, either by determining the aspect ratio of a rectangular duct that leads to a given~$\exchange$ or by using \Eqref{eqn:gap_width} as an approximation.

\subsection{The theory agrees with experimental measurements}

\begin{figure*}
	\centerline{
		\includegraphics[width=178mm]{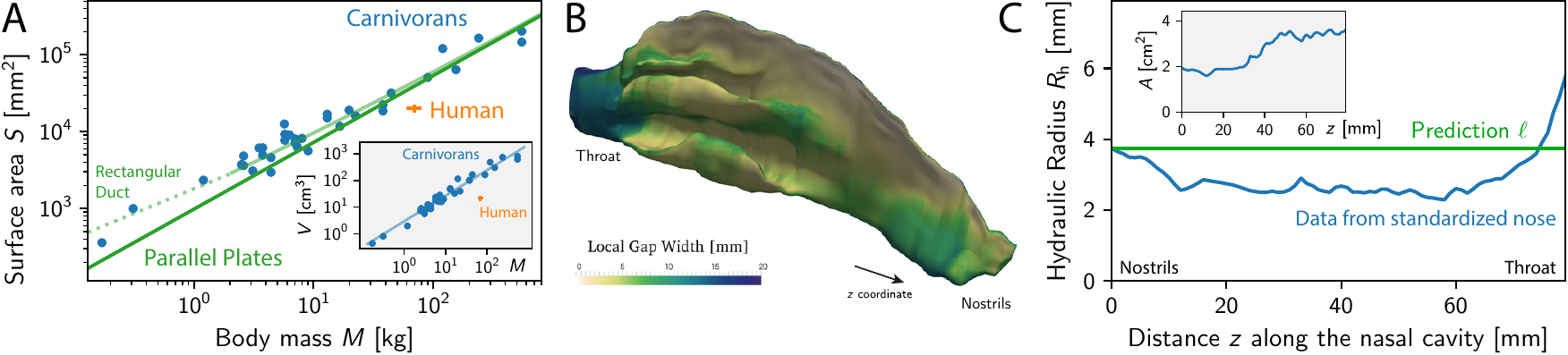}
	}
	\caption{
	Comparison of the theoretical predictions to experimental data.
	(A)~Surface area~$S$ of nasal cavities as a function of body mass~$M$.
	Shown are experimentally measured respiratory turbinal surface areas in canid and arctoid carnivorans~\cite{Green2012} (blue dots) and in humans (orange cross, bars indicate standard deviation; parameters in \tabref{tab:parameters}).
	Our predictions (green lines) follows from the optimal gap width of parallel plates given in \Eqref{eqn:gap_width} (dark green) and a numerical solution based on a rectangular geometry (light green; dashed part indicates square geometry). 
	Here, we assumed $\exchange=1$ and used the scalings given in \tabref{tab:scalings}.
	The inset shows the associated volumes~$V$ of the nasal cavities as a function of~$M$ together with the scaling given in \tabref{tab:scalings}.
	(B)~Local gap width, given by the shape diameter function~\cite{Shapira2008}, for the  standardized nasal cavity~\cite{Liu2009b}.
	(C)~Hydraulic radius~$\Rh$ along the main axis~$z$ of the standardized cavity~\cite{Liu2009b} (blue line) compared to the predicted~$\ell$ given by \Eqref{eqn:gap_width} (green line), which was calculated for $\exchange=1$ and the parameters given in \tabref{tab:parameters}.
	The inset shows the cross-sectional area~$A$ of the standardized cavity as a function of $z$.
	\label{fig:experiments}
	}
\end{figure*}

Nasal cavities described in the literature are typically narrow and exhibit little variation in gap width~\cite{Negus1954, Macrini2012}, which agrees with our theory qualitatively.
We also compare our quantitative prediction to geometric measurements of nasal cavities of canid and arctoid carnivorans that have been reconstructed \textit{in silico} from CT scans~\cite{Green2012}.

The associated scalings of the geometric parameters with body mass are summarized in \tabref{tab:scalings}.
The volumes~$V$ of the cavities and the lengths of the skulls scale isometrically, but the surface area of the cavity exhibits significant positive allometry, \ie, in heavier animals it is larger than expected from geometric scaling.

To test whether our theory can explain the observed data, we calculate the surface area of the optimal geometry of the nasal cavity as a function of the constrained parameters $D$, $A$, $L$, $Q$, and $\exchange$.
Here, $D$ is the scalar diffusivity, which is either the mass diffusivity of water vapor, $\Dvapor = \unitfrac[2.5 \cdot 10^{-5}]{m^2}{s}$ \cite{Gates2012}, or the diffusivity of heat at room temperature, $\Dheat = \unitfrac[2.2 \cdot 10^{-5}]{m^2}{s}$ \cite{Tsilingiris2008}.
Since both values are similar, we just consider the slower diffusivity, $D=\Dheat$, implying that humidification of the air will be slightly better than its heating.
The other parameters could be different for each animal, but since they have not been measured independently in all animals, we use the scaling relations with body mass for $A$, $L$, and $Q$, see \tabref{tab:scalings}.
Finally, we consider the scalar exchange efficiency $\exchange = 1$ independent of body mass, implying that temperature and humidity equilibrate with the walls to about $\unit[60]{\%}$ in all animals.
Indeed, inhaled air is typically heated to about $\unit[65]{\%}$ of body temperature~\cite{Rouadi1999} and the relative humidity is raised to about $\unit[80]{\%}$~\cite{Cole1982} in humans.

We first predict the surface area~$S$ in the simple geometry of parallel plates, where we can calculate the optimal gap width~$\ell$ explicitly using \Eqref{eqn:gap_width}.
Using the parameters given in the previous paragraph, we find $\ell \approx \unit[6]{mm} \cdot (M/\unit{kg})^{0.09 \pm 0.04}$.
The associated surface area is $S \approx 2V/\ell$ and thus scales as $S \approx \unit[9.8]{cm^2} \cdot (M/\unit{kg})^{0.87 \pm 0.07}$.
\figref{fig:experiments}A shows that this predicted scaling agrees well with the measured data. 
Consequently, our simple scaling theory correctly predicts important geometric properties of real nasal cavities.
In particular, the positive allometry of the surface area, which is observed in a wide range of animals~\cite{Owerkowicz2004,Lieberman2011}, is a direct consequence of the negative allometry of the optimal gap width.
Note that if the gap width scaled isometrically (with $M^{1/3}$), the scalar exchange efficiency~$\exchange$ would drop significantly in larger animals, because it scales as $\exchange \sim \ell^{-2}$.
Taken together, our theory shows that the allometric scaling of geometric parameters of the nasal cavity is a consequence of the physics of the airflow and the scalar exchange.

So far, we considered the idealized geometry of parallel plates, which contains unphysical periodic boundary conditions. 
For large aspect ratios, this is a good approximation to the more realistic shape of a rectangular duct, but we showed above that the gap width does not scale isometrically and the aspect shape thus varies with body mass.
In particular, smaller animals will have aspect ratios closer to unity and the scalings derived from the parallel plate model are not accurate in this case, see \figref{fig:experiments}A.
To correct this, we numerically determine the rectangular shape with a given cross-sectional area~$A$ that leads to the exchange efficiency $\exchange=1$.
The light green line in \figref{fig:experiments}A shows the associated surface area as a function of body mass, which now cannot be expressed as a simple scaling law.
Note that this correction is insignificant for large animals, which confirms that they have nasal cavities with high aspect ratio where the parallel plates model is accurate.
Conversely, there are large deviations for small animals, where the side walls contribute to the surface area significantly.
In fact, we find that the exchange efficiency exceeds $1$ for small animals even in the case of a square geometry (dashed line).
This high efficiency in small geometries thus suggests that the nasal cavities of small animals have simpler cross-sections, which has indeed been observed~\cite{Negus1954, Macrini2012}.

\begin{table}
 \caption{Typical physiological parameters for humans. 
 }
 \label{tab:parameters}
  \centering\small
  \begin{tabular*}{\hsize}{@{\extracolsep{\fill}}llr}
    Quantity & Value & Source \\ 
    \hline

    Length of nasal cavity~$L$ & $\unit[(6.5 \pm 0.7)]{cm}$ & \cite{Liu2009b} \\ 
    Volume of nasal cavity~$V$ & $\unit[(21 \pm 5)]{cm^3}$ & \cite{Liu2009b} \\ 
    Surface area~$S$ & $\unit[(200 \pm 25)]{cm^2}$ & \cite{Liu2009b} \\ 
    Cross-sectional area~$A$ & $\unit[(3 \pm 1)]{cm^2}$ & \cite{Liu2009b} \\ 
    Tidal volume~$\Vt$ & $\unit[(0.5 \pm 0.1)]{liters}$ & \cite{Herman2016, Gilbert1972} \\
    Respiratory rate~$\rateResp$ & $\unit[(15 \pm 4)]{min^{-1}}$  & \cite{Herman2016, Gilbert1972} \\
    Body mass~$M$ & $\unit[(70 \pm 10)]{kg}$  & \cite{Herman2016, Millar1986} \\
  	\hline
    Volumetric flux~$Q$ & $\unitfrac[15]{l}{min}$ & $\volFlux=2V_\mathrm t \rateResp$ \\ 
    \hline
  \end{tabular*}
\end{table}

Strikingly, one point that deviates strongly from the theoretical prediction in \figref{fig:experiments}A is for humans.
The surface area of their nasal cavity is about half of what the scaling suggests and the volume is even only about \unit[10]{\%} of the prediction (see inset). 
The data point for humans was calculated from a standardized nasal cavity, which was obtained by averaging reconstructed geometries of 30 humans~\cite{Liu2009b}, together with typical respiratory parameters given in \tabref{tab:parameters}.
To examine the geometry of human nasal cavities more closely, we compute the local gap width in the standardized nasal cavity using the shape diameter function, which gives the average distance of nearby walls at every point of the surface~\cite{Shapira2008} (\figref{fig:experiments}B), and the hydraulic radius $\Rh=2A/\Gamma$ from the cross-sectional area~$A$ and the perimeter $\Gamma$ (\figref{fig:experiments}C).
Both quantifications indicate that the gap width is remarkably constant over a large fraction of the standardized nasal cavity, while the cross-sectional area varies significantly (inset of \figref{fig:experiments}C).
However, the measured~$\Rh$ is significantly smaller than the predicted optimal gap width~$\ell \approx \unit[3.7]{mm}$, which follows from \Eqref{eqn:gap_width} together with the typical respiratory parameters summarized in \tabref{tab:parameters}.  Thus, whereas the geometry of the human
nasal cavity agrees with our qualitative result that the gap width should be constant for an efficient scalar exchange,  the quantitative prediction deviated significantly from our theory.
This is surprising since our theory worked well for all other tested mammals and this might thus hint at an exceptional behavior of the human nasal cavity.
Before we come back to this point in the discussion section, we next consider how the shape of the head constrains the nasal cavity.

\subsection{Geometric constraints imply labyrinth-like shapes}

\begin{figure}
	\centerline{
		\includegraphics[width=87mm]{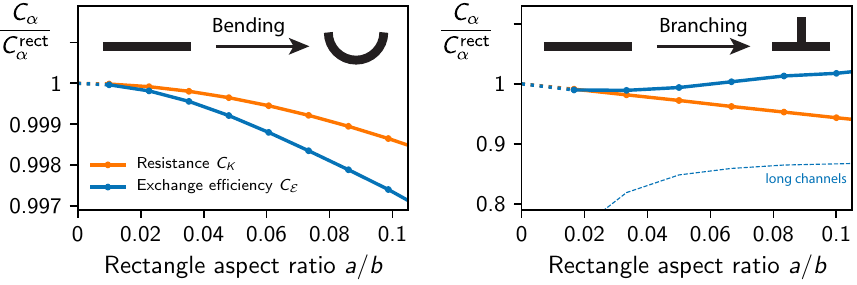}
	}
	\caption{
	Bending and branching does not affect the resistance~$\resist$ and scalar exchange efficiency~$\exchange$ significantly.
	Shown are the pre-factors $C_\resist$ (orange) and $C_\exchange$ (blue) of bent (left) and branched shapes (right) normalized to the respective values for the corresponding rectangular shape with sides $a$ and $b$ as a function of its aspect ratio $a/b$.
	$C_\resist$ follows from a numerical solution of \Eqref{eqn:flow}, while $C_\exchange$ has been calculated from \Eqref{eqn:exchange_definition} using the first 16 eigenmodes of \Eqref{eqn:scalar_eigensystem} as described in the SI.
	Parameters are given in \tabref{tab:parameters} and the dashed blue line in the right panel shows $C_\exchange$ for channels that are elongated by a factor of $100$, emphasizing the influence of the junction in longer channels.
	\label{fig:labyrinth}
	}
\end{figure}

Natural nasal cavities have a complex labyrinth-like cross-section, which does not resemble the theoretically optimally shape determined above.
This difference is likely a consequence of additional geometric constraints, since the wide rectangular cross-sections that we predict would simply not fit into the head.
However, the fact that our theory agrees well with experimental data suggests that natural nasal cavities function close to optimally. This would suggest that the bending and branching of the nasal cavity does not significantly affect the physical principles that led to the optimal gap width
 \Eqref{eqn:gap_width}. To test this hypothesis, we examine the bending and branching of the cross-section and calculate how it affects the resistance and exchange efficiency.
\figref{fig:geometries} already shows the effect of bending parallel plates into a ring shaped annulus;  the resistance~$\resist$ and exchange efficiency~$\exchange$  deviate very slightly from the parallel plates, even if the radius of curvature is only twice the gap width.
Similarly, \figref{fig:labyrinth} shows that a U-shaped cross-section has virtually identical properties to a rectangle of the same aspect ratio.
Consequently, bending the optimal cross-sectional shape in-plane does neither affect the resistance nor the exchange efficiency significantly. 
To examine the consequence of branching, we consider a T-shaped junction with three branches of equal length.
Numerical simulations indicate that both $\resist$ and $\exchange$ are affected more strongly than in the case of bending, but still only change by a few percent compared to an equivalent rectangular shape, see \figref{fig:labyrinth}.
Taken together, neither bending nor branching affects the function of the nasal cavity strongly, implying that natural shapes are close to optimal.

Another geometric constraint on the nasal cavity comes from the fact that it must connect the pharynx (and thus the lungs) to the outside world.
In humans, this forces a curved flow, which could influence the functions of the nasal cavity~\cite{Lieberman2011}, see \figref{fig:schematic}A.
In general, curved flow increases the resistance and the exchange efficiency significantly~\cite{Ito1969, Mori1967}, but in the case of the human nose the bends are localized to the connecting regions, while the main nasal cavity is rather straight.
We show in the SI that the overall function of the nose is only slightly affected by the bent geometry, consistent with numerical simulations~\cite{Nishimura2016}.
This is because the connecting regions are much wider than the main nasal cavity, so this effect is even weaker in animals that have a straighter airflow than humans.

\subsection{Gradients in the scalar exchange limit heat and humidity loss}
Up until now, we have derived the optimal geometry of the nasal cavity by focusing on the efficiency of heating and humidifying the inhaled air.
However, improving this efficiency can come at the expense of heat and water loss during exhalation.
This is because heating implies that the walls of the nasal cavity are warmer than the inhaled air, while the re-capture of heat can only occur when the walls are colder than the exhaled air.
Consequently, it is impossible to both heat the air efficiently and re-capture most of the heat during exhalation.
Such a conflicting requirement also holds for humidification and we show in the SI that small animals would lose about $\unit[1]{\%}$ of their body weight due to exhaled water each day.
To understand how to prevent this loss while still heating and humidifying inhaled air, the scalar exchange needs to be studied for both inhalation and exhalation.

To study the trade-off between heating the inhaled air and re-capturing heat during exhalation, we vary the scalar value~$\cwall$ that is prescribed at the walls of the nasal cavity.
Given that $\cwall$ is the same for inhalation and exhalation, we can calculate how the scalar transported with the air changes during these two processes.
In particular, we can define a scalar exchange efficiency~$\exchangeEx$ for exhalation analogously to $\exchangeIn$ for inhalation, given by \Eqref{eqn:exchange_definition}.
In our calculations above, we considered $\cwall = \cbody$ and $\exchangeIn=1$, which implies $\exchangeEx=0$.
Heat can be re-captured when $\cwall$ is lowered to $\cwall = \frac12(\cambient + \cbody)$, where $\cambient$ is the ambient value.
A simple analytical model presented in the SI shows that in this case $\exchangeIn=\exchangeEx\approx 0.4$.
These values can be improved to $\exchangeIn=\exchangeEx\approx 0.5$ when a linear gradient from $\cambient$ at the tip of the nose to $\cbody$ at the nasopharynx is employed.
Numerical simulations presented in the SI confirm this picture and show that the scalar exchange efficiency is actually higher than predicted by \Eqref{eqn:exchange_scaling}, because the scalar profile is typically not fully developed and entrance effects matter.
Taken together, we conclude that a gradient boundary condition, as observed in nature~\cite{Schmid1976}, can improve the recapturing of heat and humidity, at the expense of a lowered exchange efficiency during inhalation.
Note that this lower efficiency is approximately compensated by entrance effects that improve the exchange efficiency, so we expect \Eqref{eqn:gap_width} to work for nasal cavities with gradient boundary conditions and realistic lengths.

\section*{Discussion}

A critical issue for the shape of the nasal cavity are the opposing geometrical requirements for low nasal resistance and high exchange efficiency. Whereas  resistance decreases with increasing gap width, the exchange efficiency is higher when the gap is thinner. The central result of this paper is the demonstration that the optimal configuration that balances these requirements has constant gap width ~$\ell$, which we predict in \Eqref{eqn:gap_width}.
Strictly speaking, the optimal design of two parallel plates will not fit inside  the head, but our calculations show that the bending and branching of the thin duct has only a modest effect on nasal efficiency;
this suggests that the diverse morphology and labyrinth-like patterns of nasal cavities do not hinder the optimization of flow resistance and exchange efficiency.

Our theory predicts that the surface area of the nasal cavity scales as $S \sim (VQ\exchange)^{1/2}$, which implies the observed positive allometry~\cite{Owerkowicz2004, Green2012}.
This scaling explains why short-snouted animals (smaller $V$) have smaller surface area  \cite{Valkenburgh2014} and it quantifies the intuitive result that the nasal cavity can be smaller when heating and humidifying is less important (smaller $\exchange$).
The latter might explain why birds have smaller nasal cavities compared with mammals of the same weight~\cite{Owerkowicz2004,Lieberman2011}, because their relatively longer tracheas could take over part of the air-conditioning.

The scalar exchange efficiency~$\exchange$ also depends on the trade-off between conditioning the inhaled air and re-capturing the heat and moisture during exhalation.
When both processes are considered, a gradient in the boundary conditions along the nasal cavity is generally optimal, but the exact details depend on the environment and the physiological state of the animal.
Here, it will be interesting to separate evolutionary adaptations, \eg by related species living in different climates, from short-term adjustments caused by phenotypic plasticity, where for instance the gap width could be narrowed by swelling the epithelial tissue or secreting additional mucus.
Our theory predicts how such changes affect the conditioning of inhaled air and the efficiency of expelling heat with exhaled air.
It might also be used to study olfaction and the clearance of pollutants from inhaled air, which can both be described as passive scalar transport.

We also show that humans have surprisingly small nasal cavities with a reduced gap width and surface area compared to the expectations based on body mass.
In fact, the volume of the human nasal cavity is almost $\unit[90]{\%}$ smaller than expected, which then requires a narrower gap to ensure an adequate scalar exchange efficiency.
While a smaller nasal cavity likely evolved to accommodate a smaller face~\cite{Lieberman2011}, the smaller gap width implies a larger resistance to airflow.
To overcome this, the lungs could be made stronger, but this would require more energy.
Instead, humans become obligate oral breathers during heavy physical activity, which also helps dump heat~\cite{Bramble2004}.
Because the oral cavity is much wider than the nasal cavity, its resistance and scalar exchange efficiency is lower.
Consequently, heat can be dumped with the hotter exhaled air, but the inhaled air cannot be heated and humidified to the same extent as in the nasal cavity.
It will thus be interesting to compare nasal and oral breathing in humans in more detail in the future.

\subsection*{Acknowledgments}
This research was funded by the National Science Foundation through DMS-1715477 and MRSEC DMR-1420570, as well as the Simons Foundation.
D.Z. was also funded by the German Science Foundation through ZW 222/1-1.

% Bibliography
\bibliography{bibdesk}

\end{document}

\title{Supporting Information: Physical and geometric constraints explain the labyrinth-like shape of the nasal cavity}

\author{David Zwicker}
\affiliation{John A. Paulson School of Engineering and Applied Sciences, Harvard University, Cambridge, MA 02138, USA}
\affiliation{Kavli Institute for Bionano Science and Technology, Harvard University, Cambridge, MA 02138, USA}

\author{Rodolfo Ostilla-M\'{o}nico}
\affiliation{John A. Paulson School of Engineering and Applied Sciences, Harvard University, Cambridge, MA 02138, USA}
\affiliation{Kavli Institute for Bionano Science and Technology, Harvard University, Cambridge, MA 02138, USA}

\author{Daniel E. Lieberman}
\affiliation{Department of Human Evolutionary Biology, Harvard University, 11 Divinity Avenue, Cambridge, MA 02138, USA}

\author{Michael P. Brenner}
\affiliation{John A. Paulson School of Engineering and Applied Sciences, Harvard University, Cambridge, MA 02138, USA}
\affiliation{Kavli Institute for Bionano Science and Technology, Harvard University, Cambridge, MA 02138, USA}

\maketitle

\section{Pulsatile flow in the nasal cavity}

The influence of the oscillatory motion of the airflow in the nasal cavity on the resistance can be quantified by the non-dimensional Womersley number~$\Worm = \Rh (2\pi\rateResp/\viscKin)^{1/2}$~\cite{Womersley1955}.
Here, $\rateResp$ is the frequency of the oscillations and the hydraulic radius~$\Rh$ is given by the gap width~$\ell$ in our case.
Using the expression for $\ell$ given by Eq. 7 in the main text, we find
\begin{align}
	\Worm &=
	 6.89 \left(\frac{\rateResp DLA}{\viscKin \volFlux\exchange}\right)^\frac12
	 \label{eqn:scaling_womersley}
	\;.
\end{align}

Combining the scalings given in in Table~1 of the main text  with
\Eqref{eqn:scaling_womersley} implies the scaling $\Worm \approx 3.6 \cdot (M/\unit{kg})^{-0.04 \pm 0.04}$. This shows  that $\Worm$ is essentially independent of $M$, and hence similar for all animals.
For humans, using the parameters given in Table~2 in the main text, we find $\Worm \approx 1.5$. This is low enough so although pulsatility has an effect, it is insignificant~\cite{Womersley1955}.

\begin{figure}
	\centerline{
		\includegraphics[width=87mm]{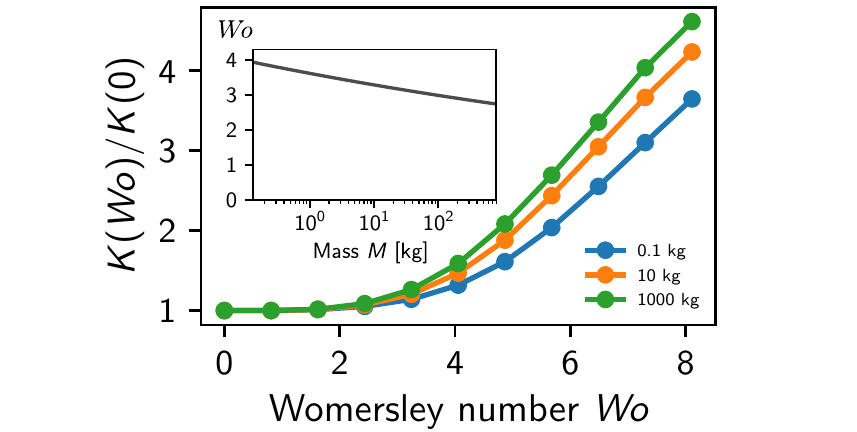}
	}
	\caption{
		Resistance~$\resist$ as a function of the Womersley number $\Worm = \Rh (2\pi\rateResp/\viscKin)^{1/2}$ for a pulsatile flow between parallel plates relative to the resistance $K(0)$ of steady flow.
		Parameters were chosen based on the scalings given in Table~1 of the main text for three different masses (colored lines).
		The inset shows that $\Worm$ does not vary much across animals.
	}
	\label{fig:pulsatility}
\end{figure}

To see this, we quantify the effects of pulsatile flow by considering flow between parallel plates driven by a sinusoidal pressure drop, $\dP(t) = \dP_0 \sin(2\pi\rateResp t)$.
Simulating this problem numerically using \comsol, we find that the resulting volumetric flux~$Q$ also exhibits a sinusoidal profile with the same frequency, $Q(t)=Q_0 \sin[2\pi\rateResp(t-\dt)]$, but lags behind the $P(t)$ by $\dt$.
Determining $Q_0$ and $\dt$ by fitting a sine function to the measured $Q(t)$, we can thus determine the resistance $K(\Worm) = \dP_0/Q_0$ associated with this pulsatile flow.
\figref{fig:pulsatility} shows that $K(\Worm)$ increases with the frequency of the flow.
However, it exceeds $K(0)$ by less than $\unit[50]{\%}$ in the range of the typical Womersley numbers between 3 and 4.
Given these relatively small effects and the fact that the influence is similar for all animals, we only consider steady flow in the main text.

\section{Resistance and scalar exchange in straight ducts}
We consider a straight duct that is oriented along the $z$-direction, so its cross-sectional domain~$\Omega$ in the $xy$-plane is independent of $z$.
For this geometry, we calculate the resistance to flow and the stationary distribution of a scalar that is advected with the flow and exchanges with the walls.

\subsection{Resistance}
The velocity profile~$\vect u$ in the duct is obtained from the stationary Navier-Stokes equations for incompressible flow,
\begin{align}
	(\vect u \cdot \Nabla)\vect u - \viscKin \Nabla^2 \vect u &= \rho^{-1} \Nabla P
&
	\Nabla \cdot \vect u = 0
\end{align}
with no-slip boundary conditions $\vect u = \vect 0$ at the walls.
For fully developed laminar flow, $\vect u = u(x,y) \, \vect e_z$ and the equations thus reduce to
\begin{align}
	\partial_x^2 u + \partial_y^2 u &= -\viscDyn^{-1} \partial_z P
	\label{eqn:flow_poisson}	
\end{align}
with the boundary condition $u=0$ on $\partial\Omega$.
Numerically, it is convenient to solve the non-dimensional Poisson equation $\partial_{\hat x}^2\hat u + \partial_{\hat y}^2\hat u = -1$ with 

scaled coordinates $\hat x = x / \xi$ and $\hat y = y / \xi$, where $\xi$ is a length scale.
Combining the associated velocity scale $(P \xi^2)/(\viscDyn L)$ with the non-dimensional area  $\hat A = \int_{\hat\Omega} \diff \hat A$ as well as the non-dimensional flux $\hat Q = \int_{\hat\Omega} \hat u \diff \hat A$, we get the dimensional variables
\begin{align}
	A &= \xi^2 \hat A
&,&&
	u &= \frac{P \xi^2}{\viscDyn L} \hat u
&\text{, and}&&
	Q &= \frac{P\xi^4}{\viscDyn L} \hat Q
	\;.
\end{align}
In particular, we can calculate the resistance~$\resist = P/Q$.
Using Eq.~2 of the main text, we thus obtain the associated pre-factor
\begin{align}
	C_\resist = \frac{\hat A^2}{\hat Q}
	\;,
\end{align}
which is related to the friction factor $2C_\resist \Rh^2 A^{-1}$.

\subsection{Scalar exchange efficiency}
\label{sec:scalar_exchange}
We next consider a passive scalar that diffuses and is advected with the velocity field~$\vect u$.
The stationary distribution $c(\vect r)$ of the scalar concentration obeys
\begin{align}
	0 &= D \Nabla^2 c - \Nabla(c \vect u)
	\label{eqn:advection_diffusion}
	\;,
\end{align}
where $D$ is the diffusivity of the scalar.
For simplicity, we here consider Dirichlet boundary conditions where the scalar concentration~$c$ is constant at the boundary.
Without loss of generality, we here consider $c = 0$ on $\partial \Omega$.

The advection-diffusion problem can be simplified since the velocity profile only has an axial component, $\vect u = u \vect e_z$.
We define the cross-sectionally averaged concentration profile $\bar c(z)= A^{-1} \int_\Omega c \,\diff x \,\diff y$ and use it to introduce the decomposition $c(\vect r) = \trans c(x,y) \bar c(z)$, where $\int_\Omega \trans c \, \diff A= 1$ by definition.
With this ansatz,  \Eqref{eqn:advection_diffusion} reduces to
\begin{align}
	D \bar c \Nabla_\perp^2 \trans c = u \trans c \, \partial_z \bar c
	\label{eqn:advection_diffusion_2}
	\;,
\end{align}
where we neglected the axial diffusion term proportional to $\partial_z^2 \bar c$, because it is dominated by the advective term, $\bar u L \gg D$.
In fact, the scalings given in Table~1 of the main text imply $\bar u L/D \approx 8.9 \cdot (M/\unit{kg})^{0.46 \pm 0.03}$, so this approximation is justified for all animals that we consider in the main text.

\begin{figure*}
	\centerline{
		\includegraphics[width=178mm]{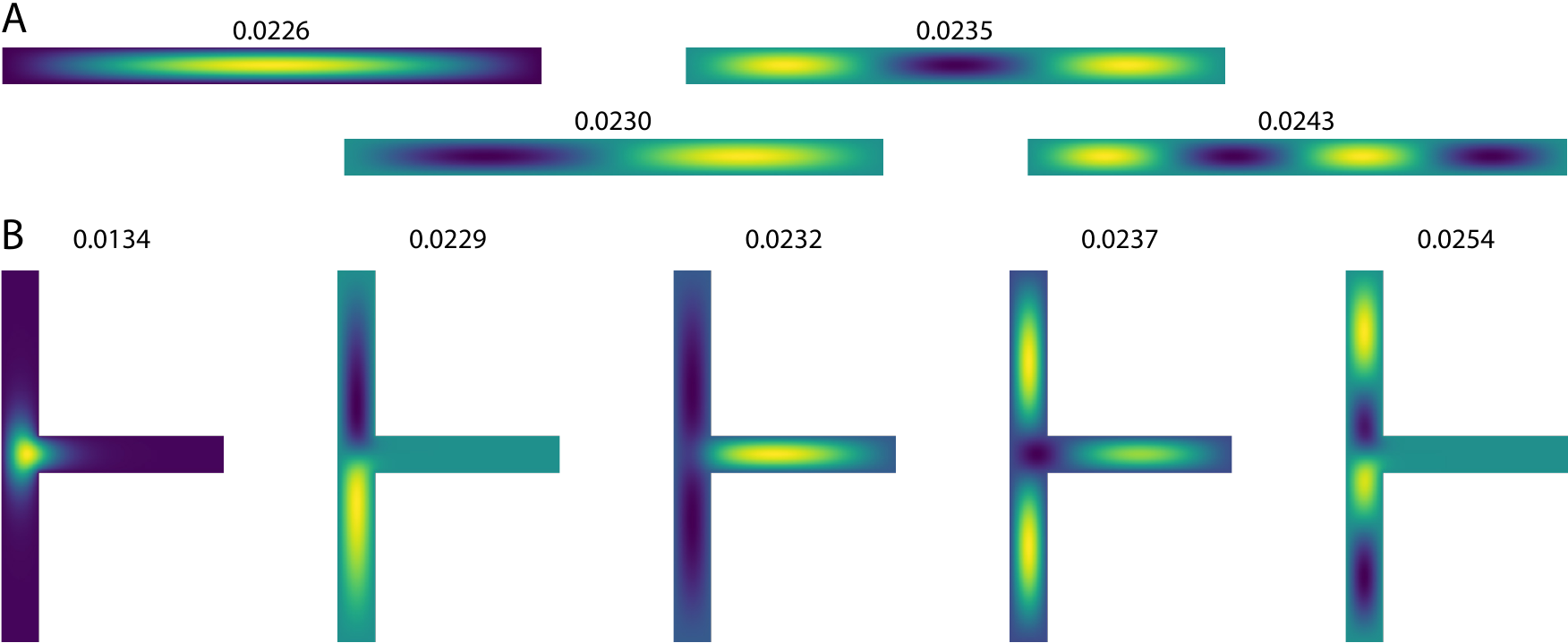}
	}
	\caption{
		Eigenfunctions~$\hat c_n$ of $\Eqref{eqn:concentration_eigenvalues_nondim}$ 
		in (A) a rectangular geometry and (B) a T-shaped geometry calculated using \comsol{}. 
		Shown are the eigenfunctions with the lowest associated eigenvalue~$\kappa_n$, which is indicated above each panel.
	}
	\label{fig:junctions_modes}
\end{figure*}

Separation of variables in \Eqref{eqn:advection_diffusion_2} implies
\begin{align}
	\frac{D \Nabla_\perp^2 \trans c}{u \trans c } = \frac{\partial_z \bar c}{\bar c } = -\Len^{-1}
	\label{eqn:scalar_separation_of_variables}
	\;,
\end{align}
where $\Len$ is a constant that needs to be determined.
The right identity in \Eqref{eqn:scalar_separation_of_variables} implies that $\bar c \sim e^{-z/\Len}$ and $\Len$ thus sets the length scale of the cross-sectionally averaged scalar profile. 
$\Len$ can be determined from the eigenvalue problem that follows from the left identity of \Eqref{eqn:scalar_separation_of_variables},
\begin{align}
	-\frac{D}{u}  \Nabla_\perp^2 \trans c  = \Len^{-1}  \trans c \;,
	\label{eqn:concentration_eigenvalues}
\end{align}
with boundary condition $\trans c = 0$ on $\partial \Omega$.
Numerically, we determine the eigenvalues from the non-dimensional problem
\begin{align}
	\partial_{\hat x}^2 \hat c_n + \partial_{\hat y}^2 \hat c_n = -\kappa_n \hat u \hat c_n
	\label{eqn:concentration_eigenvalues_nondim}
	\;,
\end{align}
where $\kappa_n$ are the eigenvalues and $\hat c_n$ are the eigenfunctions.
\figref{fig:junctions_modes} shows some typical solutions to this equation.
The decay lengths~$\lambda_n$ are then given by
\begin{align}
	\lambda_n &= \frac{P \xi^4}{\eta L D \kappa_n} = \frac{Q}{\hat QD\kappa_n}
	\;.
\end{align}
Using $\trans c_n(x,y) = \hat c_n(x/\xi,y/\xi)$, the concentration distribution of the scalar can be expressed as
\begin{align}
	c(\vect r) = \sum_n b_n \trans c_n(x, y) e^{-z/\Len_n}
	\label{eqn:scalar_series}
	\;,
\end{align}
where the coefficients~$b_n$ follow from the initial distribution $c(x,y,0)$ at the inlet.
For simplicity, we consider $c(x,y,0) = c_0$, which implies $c_0 = \sum_n b_n \trans c_n$.
Integrating both sides over $\trans c_m$, we obtain $c_0 B_m = \sum_n b_n B_{nm}$, where 
\begin{align}
	B_n &= \int_\Omega \trans c_n \, \diff x \diff y	
& \text{and} &&
	B_{nm} &= \int_\Omega \trans c_n \trans c_m \, \diff x \diff y	
	\;.
\end{align}
Consequently, $b_n = c_0 B_m B^{-1}_{nm}$, which allows for eigenfunctions~$\trans c_n$ that are not orthogonal.

The cross-sectionally averaged scalar becomes
\begin{align}
	\bar c(z) 
	&=  A^{-1}\int_\Omega \diff x\diff y \sum_n b_n \trans c_n(x, y) e^{-z/\Len_n}
\notag\\
	&=  \frac{c_0}{A} \sum_{n,m} \frac{B_nB_m}{B^{-1}_{nm}}  e^{-z/\Len_n} 
	\;.
\end{align}
Evaluating this at $z=0$ and $z=L$, we can use Eq.~(4) of the main text to give the scalar exchange efficiency as $\exchange = \ln \bar c(0) - \ln \bar c(L)$.
In particular, we obtain $\exchange =  L/\Len_1$ when the series in \Eqref{eqn:scalar_series} is dominated by the largest lengthscale~$\Len_1$.
In this case, the pre-factor for the exchange efficiency reads
\begin{align}
	C_\exchange = \frac{Q}{D\lambda_1} = \hat Q \kappa_1
	\;,
\end{align}
which is related to the Nusselt number $C_\exchange\Rh^2 A^{-1}$.

\subsection{T-shaped junctions}
The equations derived in the previous sections allow us to calculate the resistance and scalar exchange efficiency~$\exchange$ in complex geometries like the T-shaped junctions that we discuss in the main text.
The features of such geometries are apparent in the spectrum of the eigenvalue problem given in \Eqref{eqn:concentration_eigenvalues_nondim}.
\figref{fig:junctions_modes}A shows that the eigenmodes of a simple rectangular duct are regular and have all similar eigenvalues, implying that they all decay on similar length scales.
Conversely, \figref{fig:junctions_modes}B shows that the T-shaped junction has an eigenmode localized to the junction, which decays much slower than all eigenmodes that are localized to the branches.
In fact, the decay length of these eigenmodes is very similar to the ones of the rectangle, since they have the same gap width.
In contrast, the slower decay of the first mode is related to the larger effective gap width in the junction.
One consequence of the different behaviors of the modes is that the scalar exchange efficiency~$\exchange$ must be calculated using its definition given in Eq.~4 of the main text taking into account many modes.
In this case, $\exchange$ depends on the length of the channel significantly, as shown in Fig.~4 in the main text.

\section{Curved flow in human nasal cavities}

To study the effect of the bent centerline of human nasal cavities, we separate it into three parts: the external nose, the relatively straight nasal cavity, and the nasopharynx, see Fig.\ 1A of the main text.
While the nasal cavity possess the narrow geometry discussed in the main text, the other regions can be described as circular pipes of radius $\Rh \approx \unit[1]{cm}$ that are bent with an approximate radius of curvature of the centerline of $R_{\rm c} \approx \unit[3]{cm}$.

Using the parameters for humans given in Table~2 in the main text, the Reynolds number of the flow is $\Reyn = Q/(\pi\nu\Rh) \approx 500$, which implies laminar flow~\cite{Eckhardt2007}.
The effect of the curved geometry is quantified by the non-dimensional Dean number $\Dean = 2 \Reyn \, (\Rh/R_{\rm c})^{1/2} \approx 600$~\cite{Berger1983}.
For these parameters, the friction is increased by about a factor of 3 compared to a straight pipe of the same radius~\cite{Ito1969}.
Similarly, the Nusselt number, and thus the scalar exchange efficiency, increases by about a factor of 4 \cite{Mori1967}.
Consequently, the curvature has a significant effect on the resistance and the scalar exchange in both the external nose and the nasopharynx of humans, while this effect will be weaker in other animals, which typically have a less curved nasal tract~\cite{Lieberman2011}.

To see whether the increased resistance and scalar exchange due to the bent centerline matter, we next estimate the relative contribution of the external nose and the nasopharynx to the total nasal cavity.
Both regions can be described as circular pipes that are about $\unit[2]{cm}$ long.
We can thus use Eqs.\ 2 and 6 together with the data shown in Fig.\ 2 of the main text to calculate the resistance~$\resist$ and the scalar exchange efficiency~$\exchange$ of these regions.
The total resistance and exchange efficiency are then simply given by the sum of the values of the two connecting regions and the contributions from the main nasal cavity.
We find that only about $\unit[5]{\%}$ of the total resistance and $\unit[4]{\%}$ of the scalar exchange are due to the two connecting regions.
Thus, even if these contributions are multiplied by a factor of 4 due to the curved geometry, they still only comprise a small fraction and are thus negligible in the bigger picture of the scaling analysis that we present in the main text.
Taken together, the curved geometry of the external nose and the nasopharynx only contribute little to the function of the nose, mainly because they are much wider than the main nasal cavity.

\section{Gradients in boundary conditions}
We here discuss in detail the scalar exchange during inhalation and exhalation for more general boundary conditions than in the main text.

While the scalar evolves from the ambient value~$\cambient$ toward the body value~$\cbody$ during inhalation, the dynamics during exhalation are reversed, since the scalar had time to equilibrate with the body in the lungs and thus has the value $\cbody$ before entering the nasal cavity during exhalation.
To measure the scalar exchange during inhalation and exhalation, we introduce two different definitions of the exchange efficiency,
\begin{align}
	\exchange = \begin{cases}
		\ln\left(\frac{\cambient - \cbody}{\bar c(L) - \cbody}\right) & \text{inhalation} \\[8pt]
		\ln\left(\frac{\cbody - \cambient}{\bar c(L) - \cambient}\right) & \text{exhalation}  \;.
	\end{cases}
	\label{eqn:exchange_efficiency_2}
\end{align}

Here, $\bar c(L)$ denotes the cross-sectionally averaged scalar value at the outlet, \ie, at the throat for inhalation and the nostrils for exhalation.

\subsection{Simple analytical model}

\begin{figure}
	\centerline{
		\includegraphics[width=87mm]{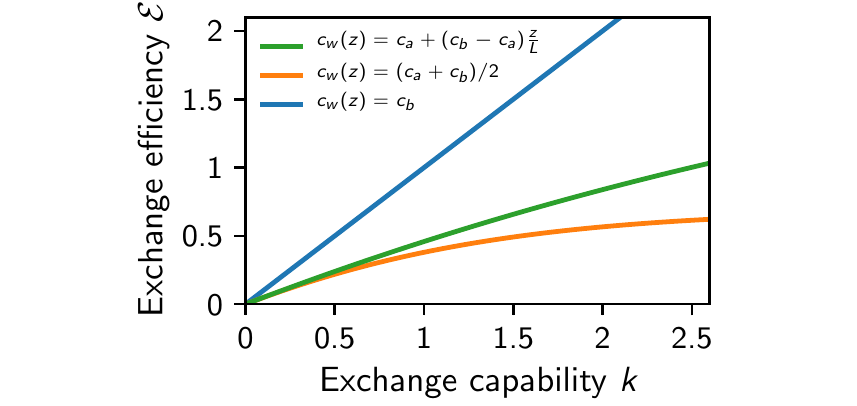}
	}
	\caption{
		Theoretical scalar exchange efficiency~$\exchange$ as a function of the exchange capability~$k$ defined in the simplified model for the scalar distribution.
		$\exchange$ is shown for the case of inhalation for three different profiles~$\cwall(z)$ of the boundary condition.
	}
	\label{fig:gradient_boundary_simple}
\end{figure}

To study complex scalar boundary conditions, we first consider a simplified model, in which the transverse variations of the scalar profile are neglected.
In this case, the change in the scalar value~$\bar c$ along the cavity is proportional to its difference to the value~$\cwall$ at the wall at this position,
\begin{align}
	\partial_t \bar c = 	\partial_z (\bar u \bar c) + \tilde k (\bar c - \cwall)
	\label{eqn:scalar_exchange_simple_ad}
	\;,
\end{align}
where we neglect the axial diffusion of $c$, similar to the treatment in \secref{sec:scalar_exchange}.
Here, $\tilde k$ quantifies the capability of the scalar to exchange with the walls, which generally depends on the scalar diffusivity, the flow, and the geometry of the channel.
In the stationary state, \Eqref{eqn:scalar_exchange_simple_ad} simplifies to
\begin{align}
	\partial_z \bar c(z) = -\frac{k}{L} \bigl[\bar c(z) - \cwall(z) \bigr]
	\label{eqn:scalar_exchange_simple}
	\;,
\end{align}
where $k = L\tilde k / \bar u$ is a non-dimensional parameter that quantifies the scalar exchange capability.

We start by analyzing the constant boundary conditions $\cwall(z) = \cbody$.
For inhalation, we solve \Eqref{eqn:scalar_exchange_simple} using the initial value $\bar c(0) = \cambient$ and find $\bar c(L) = \cbody + (\cambient - \cbody) e^{-k}$.
\Eqref{eqn:exchange_efficiency_2} then implies that $\exchange=k$ in this case.
Conversely, for exhalation, the initial value is $\bar c(0) = \cbody$ and no scalar exchange happens.
The exchange efficiency is thus zero during exhalation, so heat and humidity cannot be recaptured if $\cwall(z) = \cbody$.

One way to recapture some of the exhaled scalar would be to use an intermediate boundary value, $\cwall(z) = \frac12(\cambient + \cbody)$.
In this case we find
\begin{align}
	\bar c(L) = \frac{\cambient + \cbody}{2} + \frac{\cambient - \cbody}{2} e^{-k}
\end{align}
for inhalation.
The associated exchange efficiency reads
\begin{align}
	\exchange &= -\ln \left(\frac{1 + e^{-k}}{2} \right)
	\;,
\end{align}
and the same exchange efficiency also applies for exhalation, because of symmetry.
Consequently, the exchange efficiency during inhalation is reduced compared to the case $\cwall=\cbody$, but heat and humidity can now be recaptured during exhalation, see \figref{fig:gradient_boundary_simple}.

The simple model also allows us to discuss linear boundary conditions
\begin{align}
	\cwall(z) = c_1 + (c_2 - c_1) \frac{z}{L}
	\;,
\end{align}
where $c_1$ and $c_2$ are the values at the inlet and the outlet, respectively.
If we consider the inhalation of air in a cavity whose boundary condition changes gradually from the ambient value $c_1=\cambient$ to the body value $c_2 = \cbody$, we find that the value at the outlet reads
\begin{align}
	\bar c(L) = \cbody +  (\cambient - \cbody) \frac{1 - e^{-k}}{k}
	\;.
\end{align}
Similarly, for exhalation ($c_1 = \cbody$, $c_2 = \cambient)$, we have
\begin{align}
	\bar c(L) =  \cambient +  (\cbody - \cambient) \frac{1 - e^{-k}}{k}
	\;.
\end{align}
In both cases, the associated exchange efficiency is
\begin{align}
	\exchange = -\ln \left(\frac{1-e^{-k}}{k}\right)
\end{align}
The gradient boundary condition thus leads to higher exchange efficiency than the median condition $\cwall = \frac12 (\cambient+\cbody)$ during both inhalation and exhalation, see \figref{fig:gradient_boundary_simple}.

\subsection{Numerical simulations}

\begin{figure}
	\centerline{
		\includegraphics[width=87mm]{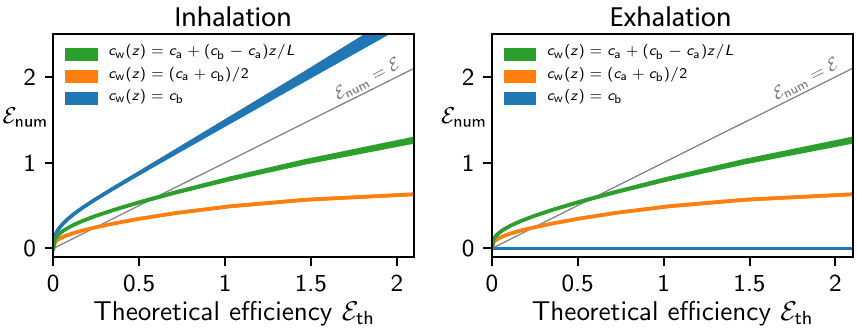}
	}
	\caption{
		The scalar boundary condition affects the exchange efficiency~$\exchange$ for inhalation (left) and exhalation (right).
		Shown is the efficiency~$\exchange_\mathrm{num}$ calculated from \Eqref{eqn:exchange_efficiency_2} assuming the ambient value~$\cambient$ outside the nose and the body value~$\cbody$ inside for different wall profiles~$\cwall(z)$.
		All values were obtained from numerical simulations of flow between parallel plates with  parameters chosen from the scalings given in Table~1 of the main text and the plate separation~$\ell$ determined from Eq.~7 with $\exchange=\exchange_\mathrm{th}$.
		The thickness of the curves indicates the standard deviation of $\exchange_\mathrm{num}$ for different body masses, $M= \unit[10^{-1} \ldots 10^3]{kg}$.
		\label{fig:scalar_gradient}
	}
\end{figure}

To test the predictions of the simple theory presented in the previous section, we performed numerical simulations using \comsol{}.
Here, we considered a parallel plate geometry with the gap width determined from Eq.~7 in the main text using a target exchange efficiency $\exchange_\mathrm{th}$ that we vary.
Solving for the stationary state, we can then simply measure $\bar c(L)$ and use the definition \Eqref{eqn:exchange_efficiency_2} to determine the real exchange efficiency~$\Enum$.

\figref{fig:scalar_gradient} shows that $\Enum$ is strongly correlated with $\exchange_\mathrm{th}$, but  $\Enum$ is consistently larger at small $\exchange_\mathrm{th}$ since the scalar profile is not fully developed.

Additionally, the qualitative shape of the curves is very similar between  \figref{fig:gradient_boundary_simple} and  \figref{fig:scalar_gradient}, suggesting that the simple model captures the essence of the scalar exchange process.
We thus conclude that a gradient in the boundary conditions is more efficient than a constant, intermediate value.
Moreover, the small width of the curves in \figref{fig:scalar_gradient} indicates that these effects are independent of the body mass of the animal and hold for all nasal cavities. 
Taken together, the exchange efficiency during inhalation is lower for gradient boundary conditions, but this allows for efficient re-capturing of heat.

\subsection{Quantifying water loss}
The mass~$m_{\rm loss}$ of water lost with exhaled vapor in a single day is given by the difference in the vapor concentration between exhaled and inhaled air multiplied by the flux~$\frac{Q}{2}$ of exhaled air,
\begin{align}
	m_{\rm loss} &= \frac{Q}{2} (c_{\rm ex} - c_{\rm in})\cdot(\unit[1]{day})
	\label{eqn:water_loss}
	\;.
\end{align}
The water loss is maximal in the extreme case of an arid environment ($c_{\rm in} \approx 0$) and no re-capturing of water during exhalation, so $c_{\rm ex} \approx \unitfrac[40]{g}{m^3}$ at a temperature of $\unit[35]{\degree C}$ and standard pressure.
Using the scalings given in Table~1 of the main text, we find that the water loss scales as $m_{\rm loss} \approx \unit[20]{g} \cdot (M/\unit{kg})^{0.78 \pm 0.02}$.
Relative to body mass~$M$, we thus have 
\begin{align}
	\frac{m_{\rm loss}}{M} &\approx 0.02 \cdot \left(\frac{M}{\unit{kg}}\right)^{-0.22 \pm 0.02}
	\;,
	\label{eqn:water_loss_relative}
\end{align}
which implies that small animals would lose a significant fraction of their body weight with exhaled humidity each day.

More typical is the case of a finite humidity in the environment, $c_{\rm in} = \cambient$, and some re-capturing of vapor in the nasal cavity.
We can estimate the latter using $c_{\rm ex} = \bar c(L)$ with $\bar c(L)$ given by \Eqref{eqn:exchange_efficiency_2}.
Using these values in \Eqref{eqn:water_loss}, we obtain
\begin{align}
	m_{\rm loss} &= \frac{Q}{2} (\cbody - \cambient) e^{-\exchange}\cdot(\unit[1]{day})
	\;.
\end{align}
The water loss is thus proportional to the difference between the body and the ambient value and is thus reduced in environments with finite humidity ($\cambient > 0$).
Moreover, a typical exchange efficiency during exhalation of $\exchange \approx 0.8$ (see \figref{fig:scalar_gradient}B) suppresses the water loss by more than two-fold.
However, the water loss still scales with body mass as given by \Eqref{eqn:water_loss_relative}, which affects smaller animals more strongly than larger ones.
Consequently, re-capturing water vapor and heat might be more important for smaller animals.

\bibliography{bibdesk}